\documentclass{epl}

\title{Nonuniversal exponents in sandpiles with stochastic particle 
number transfer}

\author{K. Jain}
\institute{
Institut 
f\"ur Theoretische Physik, Universit\"at zu K\"oln, Z\"ulpicher Strasse 77, \\
50937 K\"oln, Germany
}

\begin{document}
\def\be{\begin{equation}}
\def\ee{\end{equation}}
\def\bea{\begin{eqnarray}}
\def\eea{\end{eqnarray}}
\def\ra{\longrightarrow}
\def\l{\label}

\maketitle

\begin{abstract}
We study fixed density sandpiles in which the number 
of particles transferred to a neighbor on relaxing an active site is 
determined stochastically by a parameter $p$. Using an argument, the critical 
density at which an active-absorbing transition occurs is found exactly. 
We study the critical behavior numerically and find that the exponents 
associated with both static and time-dependent quantities vary 
continuously with $p$. 
\end{abstract}


An essential goal of the study of nonequilibrium critical phenomena is the  
identification and characterisation of the various universality classes as  
in the equilibrium case. An important prototype of nonequilibrium 
phase transitions for 
which extensive research has been done in this direction are the ones that 
occur between active states with sustained activity and 
absorbing states devoid of activity in which the system remains trapped. 
Typically the critical behavior of 
such systems falls in the directed percolation class but additional 
symmetries such as parity conservation and symmetry 
among absorbing states are known to change it \cite{aapt}.
Recent studies of active-absorbing transition in 
a class of fixed density sandpile models have shown the existence of 
a new universality class arising due to the conservation of 
global density \cite{vespignani98,alava}.

Traditional sandpile models, however, are driven-dissipative systems in 
which on exceeding a local threshold, a site becomes active (or unstable) and  
relaxes by redistributing its particles according to some rules. 
Most studies have focused on the self-organised critical (SOC) state whose 
behavior depends on the details of the model; for instance, models with 
deterministic \cite{btw} or stochastic \cite{manna} motion  
of particles, stochastic thresholds \cite{frette} 
and  ``sticky grains'' due to which an active site may or may not 
relax \cite{mohanty} have different critical exponents. 
By fixing the total density in these systems, the phase transition 
can be studied and the behavior at the critical point is found, 
under certain conditions \cite{pruessner}, to be same as that of 
the SOC state of the corresponding slowly driven 
sandpile \cite{vespignani00,christensen}.

A common feature of the sandpiles mentioned above is that the number of 
particles leaving an active site on relaxation is fixed. However, in 
experimental situations, this need not be the case. To this end, we 
introduce a class of sandpiles in which the number of particles transferred 
to a neighbor is determined stochastically. In the steady state, an 
active-absorbing transition occurs as the (conserved) 
total density of the system is tuned. We study the 
critical behavior of an order parameter 
in the steady state and its relaxation dynamics, and find that the 
associated exponents vary continuously with the stochasticity parameter.

\section{Model}
Our model is defined on a ring with $L$ sites with a non-negative integer 
$n_{i}$ of particles at each site. We call a site $i$ active if  
$n_{i}$ is greater than the threshold $n_{c}=1$ otherwise inactive. 
A configuration $C \equiv \{n_{1}, n_{2}, ..., n_{L}\}$ 
is updated in continuous time according to the following rules. 
With probability $p$, one particle hops out 
of an active site to either left or right neighbor with equal probability 
\bea
\{..., n_{i-1}, n_{i}, ...\} &\stackrel{p/2}\ra& 
\{..., n_{i-1}+1, n_{i}-1, ...\} \nonumber \\
\{..., n_{i}, n_{i+1}, ...\} &\stackrel{p/2}\ra& 
\{..., n_{i}-1, n_{i+1}+1, ...\} \;\;,\;\;n_{i} \ge 2 \;\;,
\l{pmove}
\eea
whereas with probability $q=1-p$, two particles leave an active site, one 
to the left and another to the right
\bea
\{..., n_{i-1}, n_{i}, n_{i+1}, ...\} \stackrel{q}{\ra} \{..., n_{i-1}+1, n_{i}-2, n_{i+1}+1, ...\} \;\;,\;\;n_{i} \ge 2 \;\;.
\l{qmove}
\eea
Clearly, the total density $\rho=N/L$, where $N$ is the total number of 
particles, is conserved. For any $ 0 \leq p \leq 1$, the steady state of 
this system shows a phase 
transition at the critical density $\rho_{c}$ between an absorbing phase 
in which each site has either none or one particle and an active phase with 
nonzero density $S(\rho,p)$ of active sites. 
We find numerically that the critical exponents for both the 
static and the dynamic quantities vary continuously with parameter $p$.  
Although such nonuniversal behavior has been seen in anomalous directed 
percolation which models epidemics with 
long range infections \cite{janssen} and 
in a non-local sandpile \cite{lubeck}, the model studied here has only on-site 
interactions.

Many properties of our system are known exactly in $d$ 
dimensions for $p=0$ and $1$. At $p=0$, two particles leave an active site 
and move deterministically to each of the neighbors. It is known that the 
SOC state of this model has  
density $\rho_{c}=(L-1)/L$ and the perturbations about this 
density relax in typical time $L^{z}$ with dynamic exponent $z=1$ 
\cite{dhar,ruelle}. At $p=1$, a single particle hops out 
to a neighbor which is determined stochastically. It was shown in \cite{jain} 
that in this case, an active-absorbing transition occurs at $\rho_{c}=1$ 
and due to the absence of correlations in the active phase, the activity   
$S(\rho,1) \propto \rho-\rho_{c}$. 
Further, a mapping to a two-species annihilation problem was used to  
deduce that at the critical point, the activity $S(t,\rho,1)$ at time $t$ 
decays as a power law with exponent $\theta=1/4$ until a typical time 
$L^{z}$ with $z=2$ while in the absorbing phase, it 
relaxes as a stretched exponential $e^{-(t/t_{0})^{\alpha}}$
with exponent $\alpha=1/3$. In this article, we are interested in the 
behavior of the model described above for $0 < p < 1$.

\section{Critical behavior in the steady state} 
\begin{figure}
\begin{center}
\onefigure[scale=0.333,angle=270]{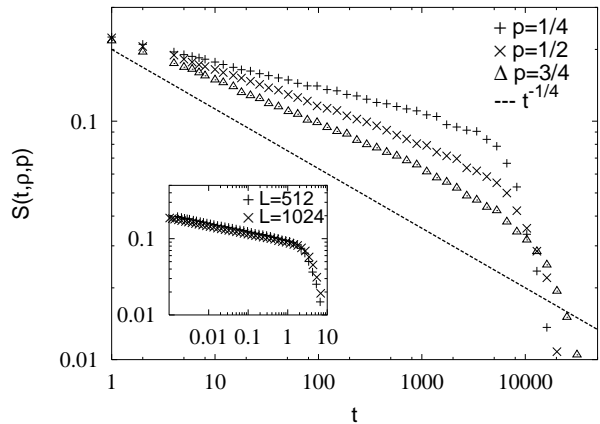}
\caption{Temporal decay of the activity $S(t,\rho,p)$  
at $\rho=1$ with $L=512$ for various values of $p$. The activity finally goes 
to zero after an initial power law regime with $p$-dependent 
exponent $\theta(p)$, $\theta(1)=1/4$. Inset: Data collapse of the 
activity in eq.(\ref{stcrit}) with $z=5/4$ at $p=1/4$.}
\label{theta}
\end{center}
\end{figure}

We first argue that the critical density $\rho_{c}=1$ for $p > 0$ and 
also give supporting numerical evidence. Since for $\rho > 1$, we 
have at least one active site, it follows that $\rho_{c} \leq 1$. 
At $\rho=1$, consider the rate of change of the 
probability of 
the configuration $C_{0}$ in which each site has one particle. In the steady 
state, since the system cannot leave this configuration, the probability of 
configurations entering it must be zero. In fact, all the 
configurations that can reach $C_{0}$ in finite number of time steps 
must have zero probability. For $p > 0$, any initial configuration can  
reach $C_{0}$ at large times by filling up its empty sites without creating 
new ones due to the single particle move (eq.(\ref{pmove})). 
Thus, $C_{0}$ is the only allowed configuration at unit density and 
since the activity $S$ is zero for $C_{0}$, the 
critical density $\rho_{c}=1$ for $0 < p \leq 1$. As shown in 
fig.~\ref{theta}  for various values of $p$, at large times, the activity 
$S(t,\rho,p)$ at $\rho=1$ decays to zero 
which confirms our assertion that $\rho_{c}=1$. 

\begin{figure}
\begin{center}
\onefigure[scale=0.333,angle=270]{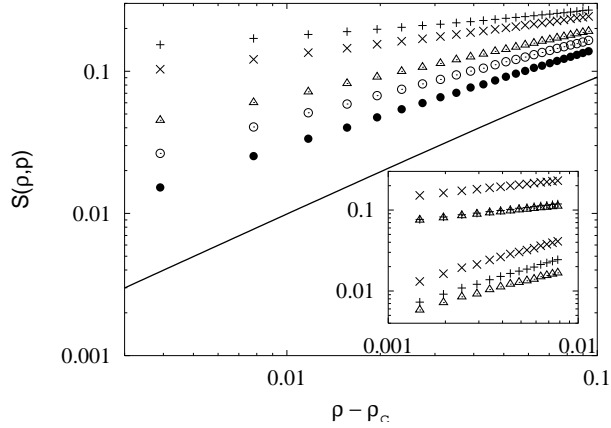}
\caption{Activity $S(\rho,p)$ vs. $\rho-\rho_{c}$ for 
$p=1/8(+)$, $1/4(\times)$, $1/2(\triangle)$, $5/8(\circ)$ and 
$3/4(\bullet)$ with $L=256$. The curve $(\rho-\rho_{c})/\rho$ shown with 
solid line is the exact expression for $S$ at $p=1$. Inset: Scaling of 
$P(0) (\triangle)$, 
$S (+)$ and $1-P(1) (\times)$ with $\rho-\rho_{c}$ for $p=1/4$ (top) 
and $p=3/4$ (bottom) with $L=2048$.} 
\label{scaling}
\end{center}
\end{figure}

Having determined the critical density, we are now in a position to 
study the critical behavior of the 
activity $S=\sum_{n \ge 2} P(n)$ where $P(n)$ is the distribution 
of the number of particles at a site in the steady state.  
Figure~\ref{scaling} shows our results obtained using 
Monte Carlo simulations for the steady state activity $S(\rho,p)$ which grows 
as a power law with deviation $\rho-\rho_{c}$ from the critical point
\be 
S(\rho,p) \propto (\rho-\rho_{c})^{\beta} \;\;,
\ee
where $\beta$ is an increasing function of $p$ with $\beta(1)=1$. 
We have verified that the above scaling behavior does not depend on the 
initial conditions. A useful check on the numerics was provided by the 
following exact relation 
\be
\rho=\frac{S+P(1)}{1-S} \;\;.\l{mass}
\ee
The above equation is just the statement of particle conservation and 
is valid for all $p$.
At $p=1$, due to the absence of empty sites in the active phase 
\cite{jain}, the normalization condition is $P(1)+S=1$. Using this in 
eq.(\ref{mass}), we immediately obtain $S \propto (\rho-1)$. For 
$p < 1$, the probability $P(0)$ is nonzero due to eq.(\ref{qmove}) so that 
in order to determine $\beta(p)$, one more nontrivial equation is required. 
The above relation, however, has the interesting consequence  
that both $P(0)$ and $1-P(1)$ also scale with $\rho-\rho_{c}$ with 
exponent $\beta(p)$ (see inset of fig.~\ref{scaling}). This scaling behavior 
can be shown for $1-P(1)$ by expanding eq.(\ref{mass}) about the critical 
density and using that $\beta(p) \leq 1$, and for $P(0)$ the desired result 
follows on using the normalization.

To derive eq.(\ref{mass}), we need to consider the equations obeyed by a 
connected two-point correlation function in the steady state; however, 
this calculation is cumbersome and lengthy, so we do not reproduce the details 
here and defer it to a future publication \cite{jainun}. 
We mention that for $p < 1$, 
our analysis also shows that the aforementioned correlation function 
vanishes, in the thermodynamic limit, for all but nearest neighbors 
indicating short range correlations in the active phase. Recall that 
at $p=1$ even these correlations are absent \cite{jain}.

To assess the importance of the correlations in the system, we 
next consider a \textit{mean field theory} which completely ignores 
the correlations. We start with the exact equations for the rate 
of change of the number distribution $P(n,t)$ at time $t$. In the 
steady state, this distribution is independent of time and we obtain
\bea
p P(n+1)+q P(n+2)+p^{\prime} 
\sum_{n^{\prime} \ge 2} P(n-1,n^{\prime}) 
&=&P(n) (1-\delta_{n,1}) + 
p^{\prime} \sum_{n^{\prime} \ge 2} P(n,n^{\prime})
,n \ge 1 \l{1pt}\\
q P(2)&=&p^{\prime} \sum_{n^{\prime} \ge 2} P(0,n^{\prime}) 
\;\;, \l{1pt0} 
\eea
where $p^{\prime}=2-p$ and $P(n,n^{\prime})$ is the joint 
distribution at two adjacent sites. In the above equations, the left 
hand side represents the elementary moves due to which $n$ particles can 
be obtained at a site, while the right hand side corresponds to the 
ways by which the particle number changes at a site having $n$ particles.

By approximating the 
joint distribution function $P(n,n^{\prime})$ in Eqs.(\ref{1pt}) and 
(\ref{1pt0}) by the product $P(n) P(n^{\prime})$, we find that the 
generating function $G(w)=\sum_{n \ge 2} P(n) w^{n}$ is given by
\be
G(w)=\frac{w^2 (1-S - (1-w) P(1))}{(w_{+}-w) (w-w_{-})} \;\;,\l{gw}
\ee 
where 
\be
w_{\pm}= \frac{1 \pm \sqrt{1+4 (1-p) (2-p) S}}{2 (2-p) S} \;\;,
\ee
with the normalisation $G(1)=S$. Due to the particle conservation 
constraint
\be
\rho= P(1)+ \frac{dG}{dw}|_{w=1}= \frac{(3- 2 p) S+(2-p) P(1)}{(2-p) (1-S)} 
\;\;.\l{rho}
\ee
As expected, this equation coincides with the exact eq.(\ref{mass}) only 
at $p=1$. 
In order to solve for $S$ as a function of $\rho$, we also need to 
determine $P(1)$. Since $G(w)$ in eq.(\ref{gw}) has a pole 
at $|w_{-}| < 1$ which 
corresponds to exponentially growing $P(n)$ with $n$, we avoid 
this unphysical solution by demanding 
that $w_{-}$ is a root of the numerator as well and obtain
$P(1)=(1-S)/(1-w_{-})$. Using this expression for $P(1)$ in 
eq.(\ref{rho}) and that the activity $S$ 
is zero at the critical density, we find $\rho_{c}=1/(2-p)$ and a 
Taylor series expansion about this $\rho_{c}$ yields 
$S \approx \rho_{c} (\rho-\rho_{c})$. 

Since these mean field predictions do not match the numerical results 
(except at $p=1$), we include some correlations by implementing 
an improved mean field theory which was successful in determining the 
fundamental 
diagram of a traffic model with parallel update \cite{goe}. We first note that 
for $p < 1$, a site with two particles can 
be emptied with probability $1-p$ but this move fills the adjacent 
(possibly empty) sites so that in the active phase, it is not possible to 
create a pair (and hence clusters) of empty sites. This is true for $p=1$ 
also since $P(0)=0$ in this case. 
Using $P(0,0)=0$ in eq.(\ref{1pt0}) and 
then approximating the rest of the joint distributions by the product of 
single site distributions, we find that while this theory correctly 
predicts the critical density to be one for all $p > 0$, it gives 
$S \propto \rho-\rho_{c}$. 
Thus, the simple mean field analyses above fail to capture the 
interesting behavior of the activity and underscores the importance of 
correlations in the system.


\begin{figure}
\begin{center}
\onefigure[scale=0.333,angle=270]{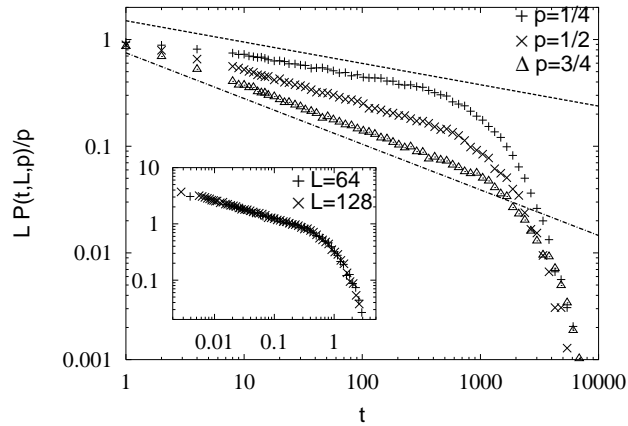}
\caption{Scaled relaxation time distribution $L P(t,L,p)/p$ at 
the critical point with $L=128$ for various values of $p$. The lines are for 
$p=1/4$ (top) and $p=3/4$ (bottom) and have slope equal to $p/(1+p)$. 
Inset: Scaling collapse of the distribution 
$P(t,L,p)$ in eq.(\ref{appcp}) with $z=3/2$ at $p=1/2$.}
\label{dynexp}
\end{center}
\end{figure}

\section{Relaxation dynamics} 
We now discuss the dynamics of the relaxation to the steady state for 
$\rho \leq \rho_{c}$. At the critical point, we 
measured the distribution $P(t,L,p)$ of the time $t$ required for the system 
to reach the steady state starting from an initial condition with a single 
empty  
site. As shown in the inset of fig.~\ref{dynexp}, the normalised distribution 
$P(t,L,p)$ is of the scaling form 
\be
P(t,L,p) \approx L^{-z} f(t/L^{z})  \;\;, \l{appcp}
\ee
where the scaling function $f(x)$ decays as $x^{-\tau}$ for $x \ll 1$ 
and exponentially for $x \gg 1$. Since the initial configuration 
can reach the critical state in one time step with probability $p/2$ 
if the empty site is adjacent to the site with two particles, and such an 
initial state occurs with probability $1/L$, it follows that 
$P(1,L,p) \sim p/L$. Using this in eq.(\ref{appcp}), we find that 
the exponents $z$ and $\tau$ are related by the scaling relation 
$z (1-\tau)=1$ with $\tau < 1$. Our simulations indicate that 
$z \approx 1+p$ which matches with the results quoted earlier for the limits 
$p=0$ and $1$, and gives $\tau \approx p/(1+p)$. 
As shown in fig.~\ref{dynexp}, the power law decay of $P(t,L,p)$ is 
consistent with this value of $\tau$. The above conjectured expression for $z$ 
was also used to obtain a data collapse for the activity to the 
following scaling form  
\be
S(t,\rho_{c},p) \approx t^{-\theta} g(t/ L^{z}) \l{stcrit} \;\;,
\ee
with $p$-dependent exponent $\theta$ (see inset of fig.~\ref{theta}).

\begin{figure}
\begin{center}
\onefigure[scale=0.333,angle=270]{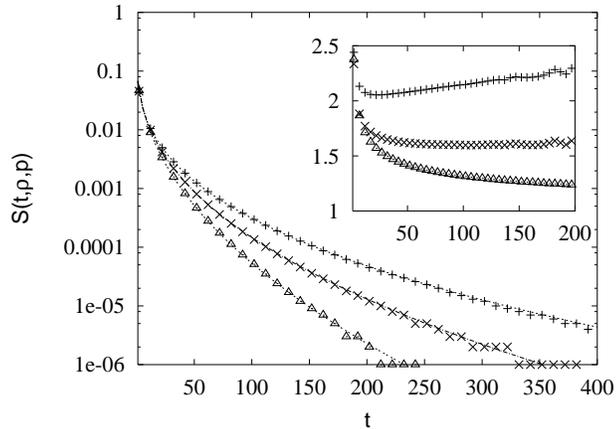}
\caption{Semilog plot to show the stretched exponential decay of the activity 
$S(t,\rho,p)$ at $\rho=1/2$ for $p=1/4 (\triangle)$, $1/2 (\times)$ and 
$3/4 (+)$ with $L=2^{20}$. Inset: Plot of $- t^{-\alpha} \ln S(t,\rho,p)$ vs. 
$t$ to determine $\alpha$ as explained in the text.}
\label{2strexp}
\end{center}
\end{figure}

We also studied the decay of the activity in the inactive phase 
with initially Poisson distributed particles. Similar to other stochastic 
sandpiles \cite{dickman}, the activity decays as a stretched 
exponential and as shown in fig.~\ref{2strexp}, we obtain a best fit to 
the form 
$S(t,\rho,p)=S_{0} \; \mbox{exp}(-(t/t_{0})^{\alpha})$ 
with $\alpha \approx 0.48, 0.42$ and $0.36$ for $p=1/4, 1/2$ and $3/4$ 
respectively. In the inset of fig.~\ref{2strexp}, we plot the equation  
\be
\frac{- \ln S}{t^{\alpha}}= 
\frac{-\ln S_{0}}{t^{\alpha}}+ \frac{1}{t_{0}^{\alpha}} \;\;, \l{reappin}
\ee
where $\alpha$ is determined by requiring that at large times, the left 
hand side approach a constant. The top curve for $p=1/4$ uses 
$\alpha=1/3$ which 
we recall is the value of the exponent $\alpha$ at $p=1$.
Since this is an increasing function at large times, the true value of 
$\alpha$ is larger than $1/3$ and in fact, a constant can be obtained with 
$\alpha \approx 0.394$. We take this as evidence to rule out 
$\alpha=1/3$ for $p < 1$. Further, since the bottom curve for $p=3/4$ 
with $\alpha \approx 0.394$ is monotonically decreasing, we conclude 
that the exponent $\alpha$ is a decreasing function of $p$ which 
agrees with the trend obtained by the best fit above. 
Thus, for $\rho \leq \rho_{c}$, the approach to the steady state is fast for 
small $p$ and slows down as $p$ increases.

We close this discussion by noting that while at $p=1$, the relaxation 
dynamics of the activity 
could be understood via a mapping to the 
well known two-species annihilation problem $A+B \rightarrow 0$ \cite{jain}, 
the corresponding reaction-diffusion system for $p < 1$ turns out to be 
more complicated involving additional reactions such as 
the ones in which both $A$ and $B$ can be created.


\section{Conclusions} We have checked that the critical behavior of our 
model is robust against perturbations. We studied this model 
with parallel update both with and without multiple 
topplings and found the same critical behavior as for the 
random sequential case discussed here. Further, a preliminary numerical 
study of the SOC state of our model supports the scaling form eq.(\ref{appcp}) 
for the relaxation time distribution $P(t,L,p)$ with the conjectured $z$. 
To conclude, we have studied several one-dimensional 
sandpiles in which the number of particles transferred on relaxing an active 
site is determined stochastically and identified 
a novel universality class of active-absorbing transitions characterised by 
nonuniversal critical exponents. An 
analytical understanding of this nonuniversal behavior is clearly 
very desirable. A promising direction seems to be a study of the SOC state 
using the operator algebra introduced in \cite{dhar}. 
While the largest eigenvalue of these operators correspond to the steady 
state, the next eigenvalue gives the dependence of the relaxation time 
on the system size $L$. For our model, we expect a $\ln L$ correction 
to the usual power law dependence of the next eigenvalue. To see this 
logarithmic function, a study of sufficiently large $L$ is 
required which would be a subject of future work. It may also be 
interesting to see if this non-universality holds in higher dimensions.

\acknowledgments The author thanks D. Dhar, M. Paczuski and 
A. Schadschneider 
for several helpful comments and J. Krug for many useful discussions and 
encouragement. This work has been supported by DFG within SFB/TR 12 
{\it Symmetries and Universality in Mesoscopic Systems.}



\end{document}